\documentstyle[preprint,aps]{revtex}
\title{Microscopic calculation of in-medium nucleon-nucleon
cross sections}
\author{G. Q. Li and R. Machleidt}
\address{Department of Physics, University of Idaho,
Moscow, ID 83843, U.S.A.}
\date{\today}
\begin{document}
\maketitle

\begin{abstract}
We derive  in-medium nucleon-nucleon (NN) cross sections
in a microscopic model. Our calculations are based upon the Bonn NN
potential and the Dirac-Brueckner approach for nuclear matter.
We consider energies up to 300 MeV (in the laboratory frame) and
densities up to twice nuclear matter density. Our results deviate
substantially from cross
section parametrizations that are commonly  used in the nuclear medium.
\end{abstract}

\section{Introduction}
An exciting topic in contemporary nuclear physics is the study of the
medium (density and/or temperature) dependence of the
properties of hadrons and hadronic processes. A well-known example is the
mass of the nucleon which decreases with  increasing  density. This
is implied by  relativistic (e.g. Walecka model \cite{qhd1,qhd2} and
Dirac-Brueckner approach \cite{shak1,shak2,mach1,brock,li1,mal1,mal2,mal3}) as
well as non-relativistic (e.g. momentum dependent Skyrme forces
\cite{brack,li2}) models. Based on effective chiral Lagrangians,
Brown and Rho \cite{brown1,brown2}
argue that the masses of other hadrons also decrease with
increasing density,  at about the same rate as the mass of the nucleon.
Recent finite density
QCD sum rule calculations also indicate that the masses of hadrons
should decrease with  increasing  density  \cite{hat,cohen1,ko1}.

Not only the static properties of hadrons (e.g. their masses) but also the
dynamical ones (e.g. two-body scattering ) are modified due to the
presence of the medium. In-medium nucleon-nucleon (NN) scattering
differs from the corresponding one in  free space, mainly due to
Pauli blocking of intermediate and final states as well as the mean field.
In conjunction with  nucleus-nucleus collisions, Faessler and
co-workers \cite{kre1,kre2,fae1} have
studied in-medium NN scattering,
based on  non-relativistic Brueckner calculations  and the Reid soft-core
potential. This approach has been applied in calculations of the
nucleus-nucleus
optical potential at low energies \cite{kre1,kre2,fae1,fae2} and
in the transport model description of nucleus-nucleus collisions at
intermediate energies \cite{aich1,aich2,li3}. It should, however, be
emphasized that,  non-relativistic many-body theory, e.g. the Brueckner
approach \cite{brue,beth,gold} and the variational method
\cite{jast,day,wir1}, is not
able to reproduce correctly the saturation properties (saturation
density and energy) of nuclear matter, when two-body forces are applied
\cite{mach1}.

The investigation of
in-medium NN scattering is of  interest
for intermediate-energy heavy-ion reactions.
Experimentally, nucleus-nucleus collisions at intermediate energies
provide a unique opportunity to form  a piece of nuclear
matter in the laboratory
with a density up to 2-3$\rho _0$ (with $\rho _0$, in the range of 0.15 to
0.19 fm$^{-3}$,
the saturation density of normal nuclear matter; in this paper we
use $\rho _0$=0.18 fm$^{-3}$) \cite{stock,grei}.
Thus it is  possible to study the properties of hadrons in
dense medium. Since this piece of dense nuclear matter exists only
for  a very short  time (typically $10^{-23}$
-$10^{-22}$ s), it is necessary
to use transport models to simulate the entire collision  process and
to deduce the properties of the intermediate stage from the known initial
conditions and  the final-state observables. At intermediate energies,
both
the mean field and the two-body collisions play an equally important role
in the dynamical evolution of the colliding system; they have to be
taken into account in the transport models on an equal footing,
together with a proper treatment of the Pauli blocking for the
in-medium two-body collisions. The Boltzmann-Uehling-Uhlenbeck (BUU)
equation \cite{bert1,mosel}
and quantum molecular dynamics (QMD)
\cite{aich3,aich4}, as well as their relativistic extensions
(RBUU and RQMD)
\cite{ko1,cass1,sorge,maru}, are promising transport models for
the description of intermediate-energy heavy-ion reactions. In
addition to the mean field, the in-medium NN cross section is also an
important ingredient of these transport models. Specifically,
in-medium total as well as differential NN cross sections are needed
by these models in dealing with the in-medium NN scattering
using a Monte Carlo method. Moreover, if these models
are used to calculate  the spectrum of particles produced in heavy-ion
collisions,
such as photons, pions and kaons,
one also needs to know the in-medium elementary particle
production cross section
\cite{mosel,cass1,ko3,li5,li6};
e.g.,  the in-medium cross section of neutron-proton ({\it np})
bremsstrahlung ($np\rightarrow np\gamma$) is needed in the
calculation of the photon production cross section in
nucleus-nucleus collisions. Up till now, most
calculations of particle production in nucleus-nucleus collisions
are based on  the free  production cross sections.

It is the purpose of this paper to calculate the elastic in-medium NN
cross sections in a microscopic way.
We base our investigation on
the Bonn meson-exchange model for the NN
interaction \cite{mach1,mach2} and the Dirac-Brueckner approach
\cite{mach1,brock,li1} for nuclear matter.

Speaking in more general terms, it is the fundamental goal of traditional
nuclear physics to describe nuclear structure and nuclear reactions in terms
of the same realistic NN interaction. This bare interaction should
be based as much as possible on theory and describe
the two-nucleon system (NN scattering as well as the deuteron) accurately.

There are two aspects to the problem. First, one
needs a realistic NN interaction which is ultimately determined by the
underlying dynamics of quarks and gluons and should in principle be
derived from quantum chromodynamics (QCD). However, due to the
nonperturbative character of QCD in the low-energy regime relevant for
nuclear physics, we are far away from a quantitative understanding
of the NN interaction in this way. On the other hand, there is a good
chance that conventional hadrons, like nucleons and mesons, remain
the relevant degrees of freedom for a wide range of low-energy nuclear physics
phenomena. In that case, the overwhelming part of the NN potential
can be constructed in terms of meson-baryon interactions. In fact,
the only quantitative NN interactions available up till now are based
upon the idea of meson-exchange; two well known examples are the Paris
potential \cite{pari1,pari2} and the
Bonn potential \cite{mach1,mach2} (see Ref. \cite{mach1} for a
comprehensive overview of the history of  meson-exchange models).

The second aspect of the problem concerns a suitable many-body
theory  that is able to deal with the bare NN interaction which
has a strong  repulsive core. The Brueckner approach
\cite{brue,beth,gold}
and variational method \cite{jast,day,wir1}
have been developed for this purpose.
However, when using two-body forces,
the Brueckner approach and the  variational method are not able
to reproduce correctly the saturation properties of
nuclear matter. Inspired by the success of the Dirac phenomenology
in intermediate-energy proton-nucleus scattering \cite{cla1,cla2,wall} and
the Walecka model (QHD) for dense nuclear matter \cite{qhd1,qhd2},
a relativistic extension of the Brueckner approach has been initiated
by Shakin and  co-workers \cite{shak1,shak2}, frequently called
the Dirac-Brueckner approach. This approach has been further developed
by Brockmann and Machleidt \cite{mach1,brock,brock1} and by ter Haar and
Malfliet \cite{mal1,mal2,mal3}. Formal aspects involved in the derivation
of the relativistic G-matrix have been discussed in detail by
Horowitz and Serot \cite{horo1,horo2}. The common feature of all
Dirac-Brueckner results is that a repulsive relativistic many-body
effect is obtained which is strongly density-dependent such that the
empirical nuclear matter saturation can be explained.
The Dirac-Brueckner approach thus provides a reasonable starting point
for pursuing the longstanding goal of self-consistently describing
nuclear matter, finite nuclei and nuclear reactions based on
the same realistic NN interaction.
The extension of this relativistic approach to the domain of
finite nuclei has been discussed by M\"uther {\it et al.}
\cite{muth} and  recently by  Brockmann and Toki \cite{brock2}, while
the investigation of nucleon-nucleus scattering
has been initiated by MacKellar {\it et al.} \cite{mac1}.

This paper is the first in a series in which we investigate systematically
in-medium NN scattering based on the Bonn potential and
the Dirac-Brueckner approach.
In this paper, we
are concerned with elastic in-medium NN scattering which
is the most important two-body process in nucleus-nucleus
collisions at incident energy below 300 MeV per nucleon.  In Section 2 we
give a brief description of the  Bonn potential and
compare  theoretical predictions with experimental data
for free-space NN scattering. The Dirac-Brueckner approach
and the predictions for nuclear matter are discussed
in Section 3. The results for the in-medium NN cross sections are
presented and discussed in Section 4. Finally we give a brief summary
and outlook in Section 5.

\section{The Bonn Model and NN Observables}

Two-nucleon scattering is described covariantly by the Bethe-Salpeter
equation \cite{beth1}. As this four-dimensional integral equation is
very difficult to solve, so-called three-dimensional reductions have
been proposed, which are more amenable to numerical solution \cite{bbs,thom}.
Among the different forms of three-dimensional reductions, the one
suggested by Thompson \cite{thom} is particularly suitable for the
relativistic
many-body problem \cite{mach1,brock}. In terms of the $R$-matrix (or
$K$-matrix) the
Thompson equation reads in the center-of-mass system \cite{mach3}
\begin{eqnarray}
R({\bf q',q})=V({\bf q',q})+{\cal P}\int {d^3k\over (2\pi )^3}V({\bf q',k})
{m^2\over E_k^2}{1\over 2E_q-2E_k}R({\bf k,q})
\end{eqnarray}
where ${\bf q,~k}$ and ${\bf q'}$ are initial, intermediate and final
relative momenta, resp., of the two scattering nucleons.
$E_q=(m^2+{\bf q}^2)^{1/2} $ with $m$
the mass of the free  nucleon. ${\cal P}$ denotes the principal value.

In the one-boson-exchange (OBE) model,
the kernel of this integral equation, $V({\bf q',q})$, is
the sum of one-particle-exchange amplitudes of certain bosons
with given mass and coupling. In the OBE Bonn model \cite{mach1},
six nonstrange
bosons with mass below 1 GeV are used; they are pseudoscalar mesons
($\pi$ and $\eta$), scalar mesons ($\delta $ and $\sigma$) and
vector mesons ($\rho $ and $\omega$). The meson-nucleon interactions
are described by  the
following Lagrangians
\begin{eqnarray}
{\cal L}_{pv}=-{f_{ps}\over m_{ps}}\overline \psi \gamma ^5\gamma ^\mu
\psi \partial _\mu \phi ^{(ps)}
\end{eqnarray}
\begin{eqnarray}
{\cal L}_{s}=g_{s}\overline \psi
\psi  \phi ^{(s)}
\end{eqnarray}
\begin{eqnarray}
{\cal L}_{v}=-g_v\overline \psi \gamma ^\mu \psi \phi _\mu ^{(v)}
-{f_v\over 4m}\overline \psi \sigma ^{\mu\nu}\psi (\partial _\mu \phi
_\nu ^{(v)}-\partial _\nu \phi _\mu ^{(v)})
\end{eqnarray}
with $\psi$ the nucleon and $\phi ^{(\alpha )}_{(\mu )}$ the meson
fields (notation and convention as in Ref. \cite{drell}). For isospin-1
mesons ($\pi ,~\delta$ and $\rho$ ), $\phi ^{(\alpha )}$ is to be
replaced by $\tau \cdot \phi ^{(\alpha )}$, with $\tau$ the usual
Pauli matrices. Note that  the pseudovector coupling is used for
pseudoscalar mesons in order to avoid unrealistically large
antiparticle contributions.

 From the above  Lagrangians, we can derive the  OBE
amplitudes; e.g., the contribution from the isoscalar-scalar meson is
given by
\begin{eqnarray}
<{\bf q'}\lambda _1'\lambda _2'|V_s^{OBE}|{\bf q}\lambda _1\lambda _2>
=-g_s^2\overline u({\bf q'},\lambda _1')u({\bf q},\lambda _1)
\overline u(-{\bf q'},\lambda _2')u(-{\bf q},\lambda _2 )
{({\cal F}_s
[({\bf q'-q})^2])^2\over ({\bf q'-q})^2+m_s^2}
\end{eqnarray}
where $\lambda _i ~(\lambda _i')$, with $i=1,2$, denotes the helicity of
the incoming (outgoing) nucleons. ${\cal F}_s[({\bf q'-q})^2]$ is a form
factor of monopole type which simulates the short-range physics
governed by quark-gluon dynamics:
\begin{eqnarray}
{\cal F}_s[({\bf q'-q})^2]={\Lambda _s^2-m_s^2\over \Lambda _s^2+({\bf
q'-q})^2}
\end{eqnarray}
with $\Lambda _s$ the cutoff mass of the isoscalar-scalar meson. The Dirac
spinors are normalized covariantly
\begin{eqnarray}
\overline u({\bf q},\lambda )u({\bf q},\lambda )=1
\end{eqnarray}
The OBE amplitudes for other mesons are  given in Refs.
\cite{mach1,mach2,mach3}.

Three sets of OBE potential parameters, denoted by  Bonn A, B
and C, have been proposed \cite{mach1}. We reprint them in Table 1.
The main difference between the three parameter sets is the cut-off parameter
for the $\pi NN$ vertex, which is 1.05, 1.2 and 1.3 GeV for Bonn A,
B and C, respectively. Consequently, the three potentials differ
in their strength of the tensor force component which is not well
constrained by present  NN data. Bonn A has the
weakest tensor force and Bonn C the strongest.
All three potentials reproduce the deuteron properties and the phase
shifts of NN scattering up to about 300 MeV
accurately (cf. Refs. \cite{mach1,brock}.

Our goal is to calculate the in-medium NN cross sections microscopically.
For this purpose it is important that the bare NN interaction  describes
the free  NN scattering cross sections correctly.
In Fig. 1,  we compare the predictions by the Bonn potentials with the
neutron-proton ($np$)
differential cross section data at 50, 129 and 212 MeV. It is
seen that all three potentials reproduce the data very well. In Fig. 2
we show the $np$ total cross sections in the energy range 50-300 MeV.
There is good  agreement between theory and experiment.

\section{Dirac-Brueckner Approach and Nuclear Matter Properties}

As mentioned in the Introduction, the essential point of the Dirac-Brueckner
approach is the use of the Dirac equation
for the description of the single-particle motion in the nuclear medium
\begin{eqnarray}
[{\bbox \alpha}\cdot
{\bf k}+\beta (m+U_S)+U_V]\tilde u({\bf k},s)=\epsilon_k
\tilde u({\bf k},s)
\end{eqnarray}
where $U_S$ is an attractive scalar field and $U_V$ the timelike
component of a repulsive vector field.
$m$ is the experimental mass of the free nucleon.
The solution of this equation is
\begin{eqnarray}
\tilde u({\bf k},s)=
\left({\tilde E_k+\tilde m\over 2\tilde m}\right)^{1/2}
\pmatrix{
1\cr
{{\bbox \sigma} \cdot {\bf k}\over \tilde E_k+\tilde m}\cr}\chi _s
\end{eqnarray}
with
$$\tilde m=m+U_S$$
$$\tilde E_k=(\tilde m^2+{\bf k}^2)^{1/2}$$
and $\chi _s$ is a Pauli spinor. Note that the in-medium Dirac spinor
is obtained from the free Dirac spinor by replacing $m$ by $\tilde m$.

The single-particle energy resulting from Eq.~(8) is
\begin{equation}
\epsilon_k = \tilde E_k + U_V \: .
\end{equation}

Similar to conventional Brueckner theory, the basic quantity in the
Dirac-Brueckner approach is a $\tilde G$-matrix which satisfies
the in-medium Thompson equation (also known as relativistic
Bethe-Goldstone equation)
\cite{mach1,brock,li1}, which reads in the nuclear matter rest frame:
\begin{eqnarray}
\tilde G({\bf q',q}|{\bf P},\tilde z)=\tilde V({\bf q',q})
+{\cal P}\int {d^3k\over (2\pi )^3}\tilde V({\bf q',k}){\tilde m^2\over
\tilde E^2_{(1/2){\bf P}+{\bf k}}}{Q({\bf k,P})\over \tilde z-
2\tilde E_{(1/2){\bf P}+{\bf k}}}\tilde G({\bf k,q}|{\bf P},\tilde z)
\end{eqnarray}
with
$$\tilde z=2\tilde E_{(1/2){\bf P}+{\bf q}}$$
and {\bf P} is the c.m. momentum of the two colliding nucleons in the nuclear
medium. Equation~(11) is density-dependent which is suppressed in our
notation.
Notice that in the energy denominator of Eq.~(11) (which is the difference
of single-particle energies of the kind Eq.~(10)),
$U_V$ drops out since it is constant.

The in-medium Thompson equation differs from the free one mainly in three
points (comparing eq. (11) with eq. (1)).
First, the Pauli operator $Q$ prevents scattering into occupied
intermediate states (`Pauli effect'). Note that this
is different from the Pauli blocking factor for the final states which is
always included in the transport models describing  nucleus-nucleus
collisions.
Second, the nucleon mean field due to the medium reduces the mass of
the nucleon and affects the energy denominator in eq. (11) which is now
density dependent, while in eq. (1) the energy denominator uses
free relativistic energies (`dispersion effect'). Finally and most importantly,
the potential used in the in-medium Thompson equation, as indicated by
the tilde, is evaluated by using the in-medium Dirac spinors of
eq. (9) instead of the free ones that are used for the $V$ in eq. (1).
This leads to the suppression of the attractive $\sigma$
exchange which increases with density.
The fact that the Dirac-Brueckner approach is able to reproduce
quantitatively the saturation properties of nuclear matter is mainly
due to this relativistic effect. This observation also implies that the
in-medium NN cross sections based on the non-relativistic Brueckner
approach lack one  important aspect
\cite{fae1,aich1,aich2},
namely, the effect which is due to the medium modification of the potential.

The scalar and vector fields of the Dirac Eq.~(8) are determined from
\begin{eqnarray}
{\tilde m\over \tilde E_i}U_S+U_V
=\sum _{j\le k_F}{\tilde m^2\over \tilde E_i\tilde E_j}
<ij|\tilde G(\tilde z)|ij-ji>
\end{eqnarray}
which is the relativistic analogue to the non-relativistic
Brueckner-Hartree-Fock definition of a single-particle potential.

Since the kernel of the in-medium Thompson equation
(eq. (11)) depends on the solution of the Dirac equation (eq. (8)),
while for the Dirac equation one needs  the
scalar and vector potentials which are related to the $\tilde G$ matrix
via Eq.~(12),
one has to carry out an iterative procedure with the goal to achieve
self-consistency of the two equations
\cite{mach1,brock,li1}: starting from reasonable
initial values for $U_S^{(0)}$ and $U_V^{(0)}$, one solves the in-medium
Thompson equation in momentum space by means of the matrix inversion
method \cite{taba} to get the $\tilde G$ matrix which leads
by means of Eq.~(12) to a new set
of values for $U_S^{(1)}$ and $U_V^{(1)}$ to be used in the next
iteration; this procedure is  continued until convergence is achieved.

The nuclear equation of state, that is the energy per nucleon,
$\cal E$/A, as a function of density, $\rho$, is obtained from the
$\tilde G$-matrix~\cite{mach1,brock}:
\begin{eqnarray}
{{\cal E}\over A}={1\over A}\sum _{i\le k_F}{m\tilde m+{\bf p}^2_i\over
\tilde E_i}+{1\over 2A}\sum _{i,j\le k_F}{\tilde m^2\over \tilde E_i
\tilde E_j}<ij|\tilde G(\tilde z)|ij-ji>-m
\end{eqnarray}

We show in Fig. 3
the energy per nucleon ${\cal E}/A$ as function of
nuclear matter density,  $\rho /\rho _0$ ($\rho _0$=0.18 fm$^{-3}$).
The solid, dashed and dotted lines are the results corresponding to the
Bonn A,  B and  C potentials, respectively. The
open rectangle  indicates the empirical region for nuclear
matter saturation. There is some difference between the
results obtained with the different potentials,
especially for the
nuclear equation of state at higher densities,
although these potentials predict the same
in free-space NN scattering.
This
difference can be traced back to differences in the strength of the tensor
force,
as reflected in the  predictions for the D-state probability, $P_D$, of
the deuteron
\cite{mach1,mach2,mach4}. The Bonn A potential, with the lowest
D-state probability ($P_D$=4.47\%), indicating a weak tensor force,
yields the  best prediction  for the
empirical values of nuclear matter saturation. More results and
discussion of the properties of (dense) nuclear matter and neutron
matter
can be found in Refs. \cite{mach1,brock,li1}.

\section{In-Medium NN Cross Sections}

In the  previous sections, we discussed briefly the Bonn potentials, their
predictions for free-space NN scattering  and the Dirac-Brueckner
approach for  nuclear matter. The potentials
describe  free-space NN scattering accurately and
nuclear matter saturation is reproduced well in the Dirac-Brueckner
approach  (especially with the Bonn A potential).

In-medium NN cross sections can be  calculated directly from the $\tilde
G$ matrix \cite{aich1,mal2,li7}. Alternatively, one  can also calculate first
 the in-medium phase shifts, which are defined in terms of the
partial-wave $\tilde G$-matrix elements
like the free-space NN phase shifts are defined in terms of
the $R$-matrix \cite{mach3} elements. From the in-medium  phase shifts, the
in-medium NN cross sections
are obtained in the usual way.

We calculate the $\tilde G$-matrix, from which we obtain the in-medium
cross sections, in the
center-of-mass (c.m.) frame of the two interacting nucleons,
i.~e., we use Eq.~(11) with ${\bf P} = 0$.
For the starting energy in Eq.~(11), we have now: $\tilde z =
2 \tilde E_q = 2 \sqrt{\tilde m^2 + q^2}$, where $q$ is related to
the kinetic energy
of the incident nucleon in the  ``laboratory system" ($E_{lab}$),
in which the other nucleon is at rest,
by: $E_{lab}=2 q^2/m$.
Thus, we consider two colliding nucleons in nuclear matter.
The Pauli projector is represented by
one Fermi sphere as in conventional nuclear matter calculations.
This
Pauli-projector, which is originally defined in the nuclear matter rest
frame, must be boosted to the c.m. frame of the two interacting nucleons. For
a detailed discussion of this and the explicit formulae, see
Refs. \cite{mal1,horo2}. In summary, for in-medium NN scattering $K$-matrix,
we use the $\tilde G$-matrix of Eq. (11) with ${\bf P}=0$,
$\tilde z = 2 \tilde E_q$,
and the Pauli-projector
$Q$ replaced by the ellipsoidal one due to Lorentz boosting.

In Fig. 4 we show the in-medium $np$ phase shifts for the $^1S_0$,
$^3P_0$  and $^3S_1$  partial-wave state as a function
of the laboratory energy. For each partial waves, we present three
different results corresponding to  density
$\rho$=0 (free-space scattering, solid lines), $\rho _0$
(dashed lines),
and $2\rho _0$ (dotted lines).
The results presented in this
figure are obtained with the Bonn A potential.
A clear decrease of the NN phase shifts
with  increasing density is observed.
This is due to the Pauli and dispersion effect as well as the
relativistic medium effects.

In Fig. 5 we show the results for the in-medium $np$ differential cross
sections
as a function of the c.m. angle
for  $E_{lab}$=50, 100, 250 and 300 MeV.
For each incident energy, we present three different
results corresponding to the medium density $\rho$=0 (free-space scattering,
solid lines), (1/2)$\rho _0$ (long-dashed lines),
$\rho _0$  (short-dashed lines) and
$2\rho _0$ (dotted lines).  The results are obtained by using the Bonn
A potential. At low incident energies (Figs. 5(a) and 5(b)),
the $np$ differential cross section always
decreases with increasing density, at both forward and backward
angles.
At high incident energies (Figs. 5(c) and 5(d)), the {\it np} differential
cross section for  forward angles  decreases when going
from $\rho$=0 to (1/2)$\rho _0$ and then increases for higher
densities. The differential
cross section at  backward angles always decreases with density.
While the free {\it np}
differential cross sections are  highly anisotropic,
the in-medium cross sections become more isotropic with increasing
density.

In Fig. 6, we  compare the in-medium {\it np} differential cross sections
as predicted by different  potentials.
The cross sections in free space are shown in Fig. 6(a), while Fig. 6(b)
and 6(c)  display them for $\rho =\rho _0 $ and $\rho =2\rho _0$,
respectively.
The incident energy is fixed at 100 MeV.
While the predictions for the np differential cross section
in free space are essentially the same,
these potentials lead
to  slightly different in-medium np differential cross sections
at higher densities (see Fig. 6(b) and 6(c)): at $\rho =\rho _0$,
the prediction by the Bonn A potential is larger than those by
Bonn B and C in all directions, while at $\rho =2\rho _0$, the
prediction by the Bonn A potential is smallest in the forward
direction and largest in the backward direction.
This difference is mainly due
to  differences in the strength of the tensor potential.
The purpose of Fig. 6 is to give some idea of the model dependence
in our approach, due to our incomplete knowledge of the nuclear force.
Fortunately, this model dependence turns out to be quite moderate.

It is also interesting to compare the
results obtained in this work with the  parametrized NN cross sections
proposed by
Cugnon {\it et al.} \cite{bert1,cugn1} which are  often used in
transport models such as BUU and QMD. This is done  in Fig. 7
for a laboratory energy of 300 MeV.
We compare our  predictions for three different densities
(using the Bonn A potential)
with the Cugnon parametrization.
It is clearly seen that, while the Cugnon parametrization
is almost isotropic,
the microscopic results still have some  anisotropy at all densities
considered. The anisotropy
in the present results decreases with
increasing density (mainly due to the decrease of the magnitude of the
$^1P_1$ phase shift).  There is also density dependence
in the microscopic differential cross section,
while  the
Cugnon parametrization is density independent.

We mention that Cugnon {\it et al}. \cite{cugn1} have parametrized the free
$pp$
cross sections. This explains the almost-isotropy in their
differential cross sections as well as the lack of the density
dependence.
Note that in the work of Cugnon {\it et al.,} \cite{cugn1}, no difference
is made between proton and neutron; thus, the $pp$ cross sections are also
used for $np$ scattering. There are, however, well-known differences
between $pp$ and $np$ cross sections which in the more accurate
microscopic calculations of the near future may be relavant. The
difference between the (in-medium) $np$ and $pp$ cross sections will be
discussed in detail in a forthcoming paper \cite{li8}. In this paper,
all ``NN cross sections" are $np$ cross sections.

The {\it np} differential cross section in  free space can be well
parametrized by the following simple expression:
\begin{equation}
{d\sigma \over d\Omega}(E_{lab}, \theta )={17.42\over 1.0+0.05(E_{lab}
^{0.7}-15.5)}exp[b(cos^2\theta+sin^2{\theta \over 7}-1.0)]
\end{equation}
with
$$
b=0.0008(E_{lab}^{0.54}-4.625) ~{\rm for} ~~E_{lab}\leq 100 {\rm MeV}
$$
and
$$
b=0.0006(36.65-E_{lab}^{0.58}) ~{\rm for} ~~E_{lab}> 100 {\rm MeV}.
$$
with $E_{lab}$ in the units of MeV.

The quality of this parametrization is demonstrated in Fig. 8, where it
is compared with the microscopic results (points) based on Bonn A at
three energies. It would be useful to parametrize the in-medium
$np$ differential cross section as well. However, the complicated
dependence of the in-medium differential cross sections on angles,
energy, and especially density makes this very difficult.
Instead, we have prepared a data file, containing
in-medium differential cross section
as a function of angle for a number of densities
and energies, from which the differential cross sections
for all densities in the range 0-3$\rho _0$ and all energies in the range 0-300
MeV can be interpolated.
This data file is available from the authors upon request.

In addition to the in-medium NN differential cross  sections which enter the
transport models to determine the direction of the outgoing nucleons,
the in-medium total NN cross sections, $\sigma _{NN}$, are also
of interest.  They provide a  criterion for  whether a pair of
nucleons will collide or not by comparing their closest distance to
$\sqrt {\sigma _{NN}/\pi }$.
We show in Fig. 9(a) the in-medium total
cross sections as function of the incident energy $E_{lab}$ and
in Fig. 9(b) as function of  density.
For completeness, we also list in Table 2 the microscopic in-medium
$np$ total cross sections as function of density and energy.
All results are obtained by
using the Bonn A potential.

It is seen that the in-medium total
cross sections decrease substantially with  increasing
density, particularly for low incident energies.
This agrees well with the findings of Ref. \cite{mal2},
but disagrees with Ref~\cite{aich1}.
In Ref.~\cite{aich1}, an enhancement of the cross section
was found below 150 MeV
and little change (as compared to the free cross section)
above 150 MeV.
The major difference between our (and Ref's.~\cite{mal2})
calculations and the ones of Ref.~\cite{aich1} is that we include
relativistic many-body effects, which are ignored in Ref.~\cite{aich1}.
Besides this, we as well as Ref.~\cite{aich1} include the Pauli and
dispersion effects (which are larger than the
relativistic effects).
The latter two effects reduce considerably the magnitude of, in
particular, the $^1S_0$ and $^3S_1$ $G$-matrix elements.
This fact is well-known since the work of Bethe and his group on
Brueckner theory in the 1960's~\cite{Day67}.
This leads to
a substantial reduction of the binding energy in nuclear matter
as well as the in-medium NN cross sections, since both are based upon
the $G$-matrix.
On the background of these well-established facts,
it is hard to understand the results of Ref.~\cite{aich1}, while
our results and the ones of Ref.~\cite{mal2} are physically
quite meaningful.

At higher energies ($\approx 300$ MeV), the medium effect becomes
smaller as compared to lower energies (cf.~Fig.~9a), but it does not
vanish. We use the continuous choice for the single-particle
potential, and so there is a dispersive effect
also for higher momenta; in addition, there are the relativistic effects.
This explains the non-negligible medium effects at intermediate energies.
Again, in this we agree with Ref.~\cite{mal2},
and disagree with Ref.~\cite{aich1} where above 150 MeV the free
cross sections were obtained. The use of the `gap-choice' for the
single-particle spectrum (i.~e., free energies above the Fermi
surface) and the omission of relativistic effects in Ref.~\cite{aich1}
may be the explanation here.

In Fig. 10 we compare the total in-medium cross sections obtained with
different  potentials. Figs. 10(a) and 10(b) correspond to the
density $\rho=\rho _0$ and 2$\rho _0$, respectively.
The solid, dashed and dotted lines
refer to  Bonn A, B and
C, respectively. By comparing
Fig. 10 with Fig. 2  we see that, although the potentials lead
to essentially the same predictions in
free space, there is some difference in the medium,
and this difference increases with density.
The prediction by the Bonn A potential is the largest at low energies and
the smallest at high energies.
Again, forturnately, the model dependence is moderate.

We compare in Fig. 11 the total in-medium cross section obtained in
the present work using the Bonn A potential
with the one used by Cugnon {\it et al}. \cite{bert1,cugn1}.
The solid, long-dashed and short-dashed lines are
the present results corresponding to the medium density $\rho=0, ~
1/2\rho _0$ and $3/2\rho _0$, respectively, while the dotted line
represents the Cugnon parametrization. It is seen that  at low  energies and
low  densities,  the Cugnon parametrization underestimates the
microscopic results, while at higher energies and higher density
it is the other way around.  Note that the Cugnon parametrization
is not density dependent, and, thus, predicts the same for all densities.

Finally,  we propose a parametrization for the total $np$ cross section as
a function of the incident energy $E_{lab}$ and density $\rho$
\begin{equation}
\sigma _{np}(E_{lab},\rho )=(31.5+0.092{\rm abs}(20.2-E_{lab}^{0.53})^{2.9})
{1.0+0.0034E_{lab}^{1.51}\rho ^2\over 1.0+21.55\rho ^{1.34}}
\end{equation}
where $E_{lab}$ and $\rho$ are in the units of MeV and fm$^{-3}$,
respectively.

We compare in Fig. 12 this parametrization (lines) with the
microscopic results (points)  based on Bonn A at four  densities.
It is seen that
the parametrization reproduces well the microscopic results
for energies and densities considered in this work.

\section{Summary and Outlook}

In this paper, we presented a microscopic derivation of elastic  in-medium
NN scattering cross sections
for energies up to 300 MeV and
densities up to 2$\rho _0$. This investigation is based
upon the Bonn NN potential and the Dirac-Brueckner approach for nuclear
matter.

The `bare' Bonn potential reproduces the
free NN scattering cross sections (differential as well as total)
accurately.
When the potential
is  applied to nuclear matter using
the Dirac-Brueckner approach,
the saturation properties
(saturation density and binding energy) are  reproduced correctly.
Thus, the Bonn model provides a good starting point for other
investigations. The major conclusions of the present microscopic
calculations are:

(1) There is strong density dependence for the in-medium cross
sections. With the increase of density,
the cross sections  decrease. This indicates that
a proper treatment of the density-dependence of the in-medium NN cross sections
is important.

(2) Our microscopic predictions  differ from the  commonly used
parametrizations of the differential  and the total
cross sections developed by Cugnon {\it et al}.
\cite{bert1,cugn1}. The Cugnon parametrization
underestimates the anisotropy of the in-medium
{\it np} differential cross sections.
In the case of  the total cross
sections, the Cugnon parametrization
either underestimates or overestimates the microscopic
results, depending on energy and  density.

At energies above 300 MeV,  inelastic channels enter the picture.
Microscopic models that also describe the inelasticity \cite{mal2}
have to be applied. These models will then also allow to calculate the
in-medium pion production.
This  is under investigation.

Acknowledgement: This work was supported  in part by the
U.S. National Science Foundation under Grant
No.   PHY-9211607,
and by the Idaho State Board of
Education. One of the authors (GQL) gratefully acknowledges  enlightening
discussions with Prof. C. M. Ko.
\pagebreak

\pagebreak
\centerline{Table 1}
\vskip 0.3cm
Parameters of the OBE Bonn potentials used in this work (reprinted from
Ref. \cite{mach1}, Appendix A, Table A.2, therein).
\vskip 0.3cm
\begin{tabular}{llllllll}
\hline\hline
\multicolumn{2}{c}{} & \multicolumn{2}{c}{Bonn A} & \multicolumn{2}{c}{Bonn B}
& \multicolumn{2}{c}{Bonn C}\\
mesons&$m_\alpha$ (MeV)&$g_\alpha ^2/4\pi$~&~$\Lambda _\alpha$ (GeV)~
&~$g_\alpha ^2/4\pi$~&~$\Lambda _\alpha$ (GeV)&~$g_\alpha ^2/4\pi$~&
{}~$\Lambda _\alpha$ (GeV)~ \\
\hline
{}~$\pi ^{a)}$~&~138.03~&~14.9~&~1.05~&~14.6~&~1.2~&~14.6~&~1.3~\\
{}~$\eta ^{a)}$~&~548.8~&~7~&~1.5~&~5~&~1.5~&~3~&~1.5~\\
{}~$\rho ^{b)}$~&~769~&~0.99~&~1.3~&~0.95~&~1.3~&~0.95~&~1.3~\\
{}~$\omega ^{b)}$~&~782.6~&~20~&~1.5~&~20~&~1.5~&~20~&~1.5~\\
{}~$\delta$~&~983~&~0.7709~&~2.0~&~3.1155~&~1.5~&~5.0742~&~1.5~\\
{}~$\sigma$~&~550~&~8.3141~&~2.0~&~8.0769~&~2.0~&~8.0279~&~1.8~\\
\hline\hline
\end{tabular}

a) $g_\alpha ={2m\over m_\alpha }f_\alpha $ for $\pi$ and $\eta$.

b) We use ${f_\rho \over g_\rho}$=6.1 and ${f_\omega \over g_\omega }$=0.

\pagebreak
\centerline{Table 2}
\vskip 0.3cm
Microscopic in-medium $np$ cross section $\sigma _{tot}$
(mb) obtained in this work with Bonn A.
$\rho _0$=0.18 fm$^{-3}$.
\vskip 0.3cm
\begin{tabular}{lllllll}
\hline\hline
 & \multicolumn{6}{c}{$E_{lab}$ (MeV)} \\
$\rho $&~50~&~100~&~150~&~200~&~250~&~300~\\
\hline
{}~0~&~164.8~&~72.18~&~49.17~&~39.57~&~34.27~&~30.85~\\
{}~$(1/2)\rho _0$~&~93.92~&~47.96~&~26.71~&~20.62~&~18.76~&~18.14~\\
{}~$\rho _0$~&~54.67~&~28.50~&~18.17~&~15.14~&~14.93~&~15.40~\\
{}~$(3/2)\rho _0$~&~37.76~&~20.93~&~15.53~&~13.91~&~14.41~&~15.26~\\
{}~$2\rho _0$~&~28.48~&~17.96~&~15.96~&~14.94~&~15.14~&~15.97~\\
\hline\hline
\end{tabular}
\pagebreak

\begin{figure}
\caption{Neutron-proton
differential cross sections at (a) 50 MeV, (b) 129 MeV
and (c) 212 MeV. The curves are predictions by the Bonn potentials.
The data at 50 MeV are from Ref. [62] (solid circles) and
Ref.  [63] (solid squares), while the data at 129 MeV
and 212 MeV are from Ref. [64]  and Ref. [65], respectively.}
\end{figure}

\begin{figure}
\caption{
Neutron-proton total cross sections in the energy range 50-300 MeV.
The curves are the predictions by the Bonn potentials. The data are
from Ref.  [66] (solid circles), Ref. [67] (solid triangles)
and Ref. [68]  (solid squares).}
\end{figure}

\begin{figure}
\caption{Energy per nucleon in nuclear matter as
obtained in the Dirac-Brueckner approach using the Bonn potentials.
The box denotes the empirical region of nuclear matter saturation.}
\end{figure}

\begin{figure}
\caption{In-medium {\it np} phase shifts for
(a) $^1S_0$, (b) $^3P_0$ and (c) $^3S_1$. The results are obtained
with the Bonn A potential.}
\end{figure}

\begin{figure}
\caption{In-medium {\it np} differential cross sections
at (a) 50 MeV, (b) 100 MeV, (c) 250 MeV and (d) 300 MeV laboratory
energy, as obtained for various densities. The Bonn A potential
is used. }
\end{figure}

\begin{figure}
\caption{In-medium {\it np} differential cross sections
at 100 MeV laboratory energy for the densities (a) $\rho $=0, (b)
$\rho =\rho _0$ and (c) $\rho =2\rho _0$. Predictions by three
differential potentials are shown.}
\end{figure}

\begin{figure}
\caption{In-medium {\it np} differential cross section at
300 MeV laboratory energy for the densities  $\rho =0$,  $\rho =\rho _0$
and  $\rho =2\rho _0$, using the Bonn A potential.
The dotted curve
is the Cugnon parametrization [33,72]}
\end{figure}

\begin{figure}
\caption{Free-space {\it np} differential cross sections at three energies.
The lines are the predictions by the parametrization, eq. (14), the squares,
circles and triangles are the microscopic results at $E_{lab}$=50, 100
and 300 MeV, respectively, as predicted by the Bonn A potential.}
\end{figure}

\begin{figure}
\caption{In-medium {\it np} total cross sections (a) as function of
the incident energy and (b) as function of  density. The
results are obtained with the Bonn A potential.}
\end{figure}

\begin{figure}
\caption{In-medium {\it np} total cross sections as predicted by
different potentials at  densities
(a) $\rho =\rho _0$ and (b) $\rho =2\rho _0$.}
\end{figure}

\begin{figure}
\caption{In-medium {\it np} total cross sections as
described by the  Cugnon parametrization [33,72] (dotted line) are
compared with the predictions by our microscopic calculation
using the Bonn A potential.}
\end{figure}

\begin{figure}
\caption{In-medium  {\it np} total  cross section as function of
incident energy at four densities.
The curves are the description by  the parametrization, eq. (15), the squares,
circles, triangles  and diamonds
are the microscopic results at $\rho =0, ~(1/2)\rho _0, ~\rho _0$ and
$2\rho _0$,
respectively, using the Bonn A potential.}
\end{figure}

\newpage

\vspace*{3cm}

The figures
are available upon request from
\begin{center}
{\sc gqli@tamcomp.bitnet}
\end{center}
Please, include your mailing address with your request.

\end{document}